# A catalogue of Type 2 active galactic nuclei in the 6dF Galaxy Survey

Sruthi Suresh[1,2]★, Wei Jeat Hon[1,3], Rachel L. Webster[1,2], Christian Wolf[3,4] and Christopher A. Onken[3,4]

[1]*School of Physics, University of Melbourne, Parkville, VIC, 3010, Australia*
[2]*ARC Centre of Excellence for All Sky Astrophysics in 3 Dimensions (ASTRO 3D), Australia*
[3]*Research School of Astronomy and Astrophysics (RSAA), Australian National University, Canberra, ACT 2611, Australia*
[4]*Centre for Gravitational Astrophysics, Australian National University, Building 38 Science Road, Acton, ACT 2601, Australia*

**ABSTRACT**
Active galactic nuclei (AGNs) are the compact, energetic central regions of galaxies, powered by supermassive black holes that accrete surrounding gas and dust. Their optical spectra can be identified by strong emission-line signatures (broad and/or narrow lines). Those showing only narrow lines are classified as 'Type 2' AGN. Extensive surveys like SDSS cover AGN in the Northern Sky, but the equivalent coverage in the Southern Sky remains limited. We address this by presenting a new catalogue of Type 2 AGN from the 6dF Galaxy Survey (6dFGS), which has 136 304 spectra covering $\sim$ 17 000 deg$^2$, mainly of low-redshift galaxies. We use a median absolute deviation cut on the continuum-fitted spectrum to select emission line galaxies. AGN were identified by fitting the 6dFGS spectra with a modified Python QSO fitting tool (PyQSOFit). All selected spectra were visually inspected and corrected for fitting errors where necessary. 10 492 narrow emission line galaxies were identified in 6dFGS, including $\sim$5000 Type 2 AGN classified from Baldwin–Phillips–Terlevich (BPT) diagram. They have a median redshift of $z \sim 0.032$ and a median [O III] luminosity of $\log(L_{[\text{O\,III}]}/\text{ergs s}^{-1}) \approx 40.04$.

**Key words:** techniques: spectroscopic – catalogues – galaxies: active – galaxies: nuclei – (*galaxies*:) quasars: emission lines – galaxies: Seyfert.

## 1 INTRODUCTION

The first observational hints of active galactic nuclei (AGNs) emerged in early 20th century when spectroscopic studies of the brightest 'Spiral Nebulae' revealed unusual features (Fath 1911). One peculiar source, now known as NGC 1068, exhibited six emission lines instead of a continuous spectrum, indicating a different energetic processes occurring at the galactic core. A few decades later, Seyfert (1943) found strong ionizing emission lines in a few more spiral galaxies. Since then, the study of AGN has advanced, and distinct features have been identified. An AGN is now defined as the galaxy's compact, energetic, central engine powered by a supermassive black hole. These nuclei are extremely luminous objects, emitting a broad range of wavelengths across the electromagnetic spectrum. The formation of these supermassive black holes is still poorly understood. The interaction with the host galaxy, AGN triggering, and the activity cycle of an AGN are some of the intriguing questions yet to be understood.

The most common method used to identify an AGN is the presence of strong emission lines in the optical spectrum (Kauffmann et al. 2003; Brinchmann et al. 2004; Kewley et al. 2006; Stasińska et al. 2006). The identified AGNs are characterized by features based on the presence of broad and/or narrow emission lines. Broad emission lines arise in the central engine due to the presence of high-velocity gas, causing Doppler broadening of $\gtrsim$1000 km s$^{-1}$ (Hao et al. 2005; Stern & Laor 2012). This type of galaxy is called a 'Type 1' AGN. Narrow emission lines, on the contrary, are formed outside the nuclear region by lower velocity gas clouds, with a Doppler broadening width of few hundreds of km s$^{-1}$ (roughly $\lesssim$1000 km s$^{-1}$) (Osterbrock 1993; Padovani et al. 2017). Due to the orientation of these galaxies, the broad-line region is obscured by dust and/or gas in a torus around the accretion disc and such galaxies are classified as 'Type 2' AGN (Antonucci 1993; Villarroel & Korn 2014). Although the absence of broad lines in a Type 2 AGN is commonly explained by obscuration by a dusty torus, an alternative explanation might be the existence of 'true' Type 2 AGN which may not have a broad-line emitting region (Tran 2001; Panessa & Bassani 2002; Elitzur & Netzer 2016).

Optical surveys such as the Sloan Digital Sky Survey (SDSS; Blanton et al. 2017) provide researchers with ample data to construct catalogues of specific objects. With the abundance of telescopes located in the Northern hemisphere, rich AGN data sets have been constructed in the Northern Sky. Some of the surveys that identify the AGN are: in the optical (Schneider et al. 2010), optical-infrared (Zaw, Chen & Farrar 2019; Chen et al. 2022, hereafter C22 and Z19 respectively), X-ray (Alexander et al. 2003; Panessa et al. 2014; Belvedersky et al. 2025) or in the Radio (Becker, White & Helfand 1995; Gizani & Leahy 2003).

The Six-Degree Field Galaxy survey (6dFGS; Jones et al. 2004, 2009, hereafter J04 and J09, respectively) provides a complete data set of galaxies with optical spectra across a large area of the Southern

★ E-mail: sruthis@student.unimelb.edu.au

Sky. The 6dFGS catalogue is similar to the spectroscopic component of SDSS for the Northern Sky, although the spectral resolution of the former is poor compared to the latter. This difference is due to the distinct scientific goals behind these surveys. The primary objective of 6dFGS was to map the nearby galaxies in the Southern Sky and thus it focused on obtaining galaxy dynamics, peculiar velocities and redshifts. However, a comprehensive and complete optical AGN catalogue for the Southern Sky has not been produced from 6dFGS to date. The gap in AGN studies in the Southern Sky would be keenly felt in the era of the Legacy Survey of Space and Time (LSST; NSF-DOE Vera C. Rubin Observatory 2025). LSST at the Vera C. Rubin observatory will image every part of the Southern Sky about once a week for ten years and provide the variability light curves for AGN (Ivezić et al. 2019). However, fully exploiting LSST requires the identification of complete AGN catalogues in that part of the sky.

The main motivation of our work is to construct an extensive catalogue of AGN using the magnitude-limited 6dFGS optical spectral data. We use PyQSOFit (Guo, Shen & Wang 2018), a PYTHON tool adapted from Yue Shen's IDL code, to fit spectra and measure emission lines. It includes host templates, Fe II models, and reddening corrections, and supports Monte Carlo uncertainty estimation used in AGN/quasar studies (Panda & Śniegowska 2024; Ren et al. 2024). We have constructed a catalogue of Type 2 Narrow Line Galaxies (NLGs), which are then classified using the Baldwin–Phillips–Terlevich (BPT) diagram (Baldwin, Phillips & Terlevich 1981, hereafter B81). The BPT diagram effectively sorts emission-line galaxies into different classes, based on the ionization properties revealed by emission line ratios. This allows us to classify the Type 2 NLGs into star-forming (SF), composite, or AGN (Veron 1981; Keel 1983; Veilleux & Osterbrock 1987). Empirical boundaries by Kewley et al. (2001) and Kauffmann et al. (2003) further refine this classification.

With this work, we achieve the goal of making a complete catalogue of Type 1 (Hon, Webster & Wolf 2025, hereafter H25) and Type 2 (this work) AGN. All sources in these catalogues are verified by visually inspecting the fitted optical spectra. The source selection criteria for both catalogues are similar. We have aimed to maintain consistency in the methodology to facilitate combining both data sets, if needed, for scientific analysis.

Similar catalogues were published by C22 and Z19, who have also identified AGN using the 6dFGS data. A major difference lies in providing the best-fitting parameters by the inspection of individual optical spectra. A more detailed comparison of both catalogues is discussed in Section 6.1.

This paper is organized as follows. In Section 2, we provide detailed information about our parent sample, 6dFGS. Section 3 explains the statistical methods used to filter for emission line galaxies. A flowchart (Fig. 8) outlines the methods deployed in this work to identify Type 2 AGN. Section 4 briefly explains the method used to classify emission line galaxies. Section 5 discusses the properties of the catalogue, while Section 6 analyses the catalogue to find various trends in the data. We summarize our work in Section 7, briefly mentioning planned future work.

We adopt a flat $\Lambda$CDM cosmology with $\Omega_\Lambda = 0.7$ and $H_0 = 70$ km s$^{-1}$ Mpc$^{-1}$. Optical magnitudes are in the AB system and IR magnitudes are in the Vega system. Throughout this paper, references to the [O III] line specifically denote [O III]$\lambda$5007 unless stated otherwise. Likewise, [N II] refers to the [N II]$\lambda$6583 line.

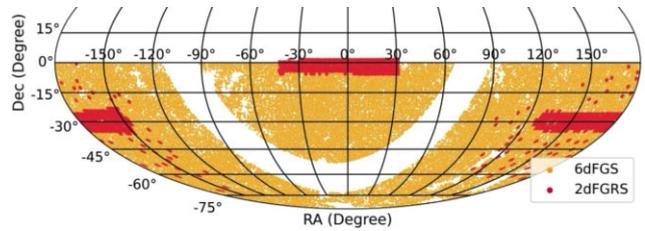

**Figure 1.** The sky map of 6dFGS (yellow) and 2dFGRS (red) sources. The former survey covers approximately ten times more sky than the latter.

## 2 DATA SAMPLE

The parent sample of our AGN catalogue is the 6dF Galaxy Survey (6dFGS; J09), which was aimed at obtaining a complete redshift catalogue for a flux-limited sample of nearby galaxies in the Southern hemisphere, excluding regions close to the Galactic plane. Observations for the survey took place between May 2001 and January 2006 using the 6dF fibre-fed multi-object spectrograph mounted on the United Kingdom Schmidt Telescope (UKST). The survey targeted regions covering approximately 17 000 square degrees of the Southern Sky, located more than 10 degrees from the Galactic plane. This area is roughly ten times larger than that covered by the 2dF Galaxy Redshift Survey (2dFGRS; Colless et al. 2001) and more than twice the spectroscopic coverage of the SDSS Data Release 7. The sky distribution of 6dFGS and 2dFGRS sources is shown in Fig. 1.

6dFGS primarily focuses on near-infrared (NIR) and optical selection criteria. NIR magnitudes are derived from the 2MASS Extended Source Catalog (XSC) and represent total extrapolated magnitudes (Jarrett et al. 2000). The $K$-band sample, with a completeness limit of $K = 12.65$ mag, constitutes the primary target selection. Supplementary targets were included to ensure completeness in other wavebands, namely $J$, $H$, $b_J$, and $r_F$, with limits of 12.95, 13.75, 15.60, and 16.75 mag, respectively. Various smaller samples, drawn from different catalogues and wavelengths, were also incorporated to complete the allocations of targets for spectroscopy.

From 136 304 observed spectra, J09 obtained 110 256 extragalactic redshifts and, together with a smaller number of known redshifts, compiled a catalogue of 125 071 galaxies. The final sample comprises of 113 747 flux limited (main targets) while 22 557 (Auxiliary sources) are selected from various other catalogues (see Table 1). Approximately, 8 per cent of these targets already had redshifts from prior surveys such as ZCAT (Huchra et al. 1992), 2dFGRS (Colless et al. 2001), or the SDSS (SDSS DR7, Abazajian et al. 2009). 6dFGS obtained spectra from 136 304 source observations, resulting in 126 754 unique redshifts of varying quality. The median redshift of this sample is 0.053.

### 2.1 Redshift measurements

The main goal of 6dFGS was to measure redshifts and study cosmic structure across a hemisphere. For absorption line galaxies, 6dFGS spectra were cross-correlated with a matching template spectra. The high signal-to-noise template spectra include a stellar spectra (K-type giants), synthetic or composite galaxy templates. The cross-correlation function determines the velocity shift that aligns with the observed spectrum, deriving redshift. For spectra dominated by emission lines, redshift was determined by identifying prominent features such as H$\beta$, [O III], and H$\alpha$. A Gaussian fit to the emission lines provides the offset from their rest-frame wavelengths, which is then used to derive the redshift.

**Table 1.** List of subsamples in the overall 6dFGS sample.

| PROG_ID | Sample name | Count |
|---|---|---|
| | Flux limited | |
| 1 | 2MASS Kext $\leq$ 13.00 | 97 020 |
| 3 | 2MASS Hext $\leq$ 13.05 | 2021 |
| 4 | 2MASS Jext $\leq$ 13.75 | 1284 |
| 5 | DENIS J $\leq$ 14.00 | 629 |
| 6 | DENIS I $\leq$ 14.85 | 504 |
| 7 | SuperCOSMOS $r_F \leq$ 15.60 | 5773 |
| 8 | SuperCOSMOS $b_J \leq$ 16.75 | 6516 |
| | Auxiliary | |
| 78 | UKST | 271 |
| 90 | Shapeley supercluster | 630 |
| 109 | Horologium Sample | 469 |
| 113 | ROSAT All-Sky Survey | 1961 |
| 116 | 2MASS Red AGN | 1141 |
| 119 | HIPASS | 439 |
| 125 | SUMSS/NVSS radio | 2978 |
| 126 | IRAD FSC | 5994 |
| 129 | Hamburg-ESO QSOs | 2006 |
| 130 | NRAO-VLA QSOs | 2673 |
| 999 | Unassigned | 3995 |

Redshift quality (Q) in the 6dFGS was assessed visually on a scale of 1 to 6, with Q = 1 indicating unusable measurements and Q = 6 assigned to stars and confirmed Galactic sources. For the galaxy analysis, J09 recommend using only Q = 3 and Q = 4. Hence, our parent sample comprises 117 610 sources.

### 2.2 Six-degree spectrograph

The Six-Degree Field multi-object fibre spectroscopy facility (6dF) at the UKST is located at the Siding Spring Observatory near Coonabarabran in Australia. The instrument has three main components including two interchangeable 6dF field plates, where one can be configured while the other is used for observing. 6dF can observe up to 150 spectra simultaneously, across the 5.7° diameter field. There are some observational restrictions in the 6dFGS which are mentioned in J04.

One of the main observational restrictions, relevant to this work, is the two-grating set-up. To observe sources in the wavelength range 4000–7500 Å, each field had to be observed separately with two grating set-ups, namely V-arm and R-arm. Before October 2002, V-arm spectra spanned 4000–5600Å, while R-arm spectra spanned 5500–8400 Å. Subsequent observations (around 80 per cent) feature V-arm spectra spanning 3900–5600 Å and R-arm spectra spanning 5300–7500 Å. V-arm spectra offer a resolution of 800–960 (5–6 Å) at the full wavelength at half-maximum (FWHM), while R-arm spectra have a resolution of 645–716 (9–12 Å) (see J09 for more details).

At low redshift, the V-arm spectrum covers the H$\beta$ and [O III] lines, whereas the R-arm spectrum encompasses the H$\alpha$ and [N II] lines. To apply the BPT diagram in the analysis of the AGN, it is thus essential to have both V-arm and R-arm spectra. However, this also imposes a strict redshift cut to our AGN sample at $\sim z = 0.27$.

Fig. 2 shows the redshift distribution in the 6dFGS main sample along with the redshift distribution of Type 2 Emission line galaxies identified in this work. 6dFGS auxiliary sources have redshifts up to 3, but Fig. 2 is truncated to our upper limit of 0.27, which excludes 5.1 per cent of 6dFGS.

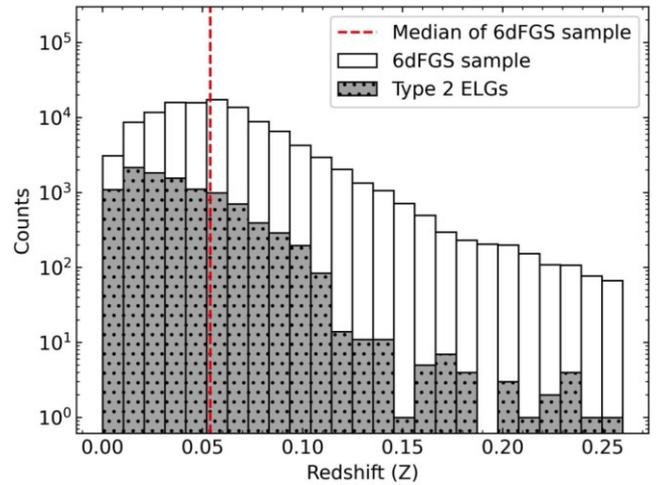

**Figure 2.** Histogram of redshifts in the overall 6dFGS sample versus Type 2 AGN (this work). The dashed line indicates the median value of the 6dFGS sample. The upper limit is imposed by our requirement to retain H$\alpha$ and [N II], whereas the 6dFGS sample includes $\sim$6000 sources at higher redshift.

### 2.3 Challenges and cross-talk in 6dFGS

As mentioned in the previous section, any 6dFGS optical spectrum is composed of two separately observed parts. The main motivation for the 6dF galaxy survey is to measure redshifts for as many galaxies as possible. Thus, the spectra are not flux-calibrated and have a signal-to-noise ratio that is good enough for redshift determination but not ideal for a physical analysis of galaxies. In the following, we list some of the challenges and limitations in 6dFGS, that have also been mentioned in J04:

(i) Some of the sources in 6dFGS are observed in marginal conditions which leads to a low S/N ratio. Sources with poor data were re-observed to get better results.
(ii) Some fibres have poor throughput due to technical challenges including misalignment.
(iii) Many sources in 6dFGS are missing useful red data as they suffer increasing levels of fringing towards the longer wavelength.
(iv) Spurious spectral features occur at various wavelengths: some sources have spurious peaks at 4440 Å in the V grating and at 6430 and 6570 Å in the R grating, which cover a 10-pixel-wide region.
(v) The well-known strong night-sky emission line at 5577 Å may fall on the [O III] or H$\beta$ line, depending on redshift.
(vi) The separately observed V and R arms of the spectrum have different noise levels and sometimes different continuum levels. Automated fitting fails in these cases and manual tuning of the fitting procedure is required.
(vii) A major issue with 6dFGS is fibre cross-talk: the spectra of bright sources can contaminate those of faint sources whose light is fed through neighbouring fibres on the slit. The fainter source might then show a second set of spectral lines, which really belongs to the brighter source in the neighbouring fibre. In some sources the cross-talk is so strong that they need to be removed from the parent sample. An incomplete list of cross-talk affected galaxies is shown on the 6dFGS website.[1] In others, the cross-talk signal can be identified and safely ignored. Such sources are flagged in the catalogue and listed in the Table 2. An example of cross-talk is shown in Fig. 3.

**Table 2.** List of cross-talk spectra identified in this work. The columns Name and Spec ID list the 6dFGS source name and its corresponding spectra ID of the affected spectra. The Cross-talk spec ID column indicates the spectral ID of the interfering source. The final column lists the spectral IDs of the sources that were retained for analysis in this study. The asterisk (∗) in the column name indicate sources already identified and listed in the 6dFGS website.

| Name | Spec ID | Cross-talk spec ID | Spec ID, if included |
|---|---|---|---|
| g0034272-031219∗ | 3161 | 3160 | |
| g0109141-722940∗ | 4609 | 4610 | 4609 |
| g0318456-061646 | 18553 | 18552 | 18553 |
| g0409123-232321 | 23934 | 23935 | 23934 |
| g0419223-085443∗ | 24754 | 24755 | |
| g0503542-395532 | 28478 | 28477 | 28478 |
| g0501038-244649∗ | 29168 | 29169 | 29168 |
| g0510245-235819 | 29337 | 29338 | 29337 |
| g0511030-092230 | 29809 | 29808 | 29809 |
| g0513026-091238 | 29810 | 29808 | |
| g0556523-052309 | 34423 | 34425 | 34423 |
| g0651157-440329 | 38492 | 38491 | 38492 |
| g0716300-520511 | 39360 | 39361 | 39360 |
| g0742250-553943 | 40174 | 40173 | 40174 |
| g0749090-561457 | 40190 | 40191 | |
| g0823036-080905 | 42004 | 42005 | |
| g1000193-411934 | 47209 | 47208 | |
| g1000031-293637 | 48544 | 48545 | |
| g0958137-292451 | 48545 | 48544 | 48545 |
| g1004261-282639 | 48553 | 48552 | |
| g1328344-030745 | 66254 | 66253 | |
| g1334348-245326∗ | 66724 | 66723 | |
| g0557440-211655 | 125919 | 125920 | 125919 |
| g0655590-404912 | 126131 | 126133 | 126131 |

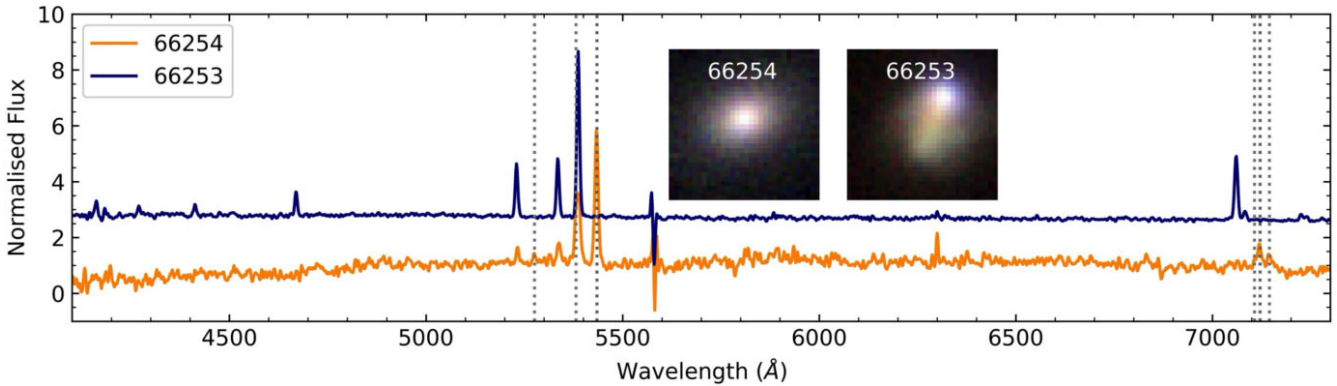

**Figure 3.** An example of cross-talk in the *V*-arm is shown in the figure. The emission lines of the galaxy with the Spec ID 66 253 affect the galaxy with Spec ID 66 254. The [O III]λ4959 line of Spec ID 66 254 is affected by [O III]λ5007 of Spec ID 66 253. PanSTARRS image cutouts of the galaxies are included as inset images. The dotted line indicates the wavelength of Hβ, [O III]λλ4959,5007, Hα, and [N II]λλ6548,6583 of the galaxy with Spec ID 66 254.

## 3 SPECTRUM PRE-PROCESSING AND FITTING

As described in Section 2, the parent sample, 6dFGS aims to study the redshifts of all types of galaxies, necessitating a filtering to isolate AGN candidates. With a vast data set comprising 117 610 spectra, manually fitting each spectrum or visually inspecting all is impractical. Employing statistical analysis becomes imperative to efficiently process such large volumes of data. The following subsections explain the pre-processing techniques employed in this work.

### 3.1 Spectrum cleaning and continuum subtraction

Like all the optical spectrum, 6dFGS also has some unwanted noisy artefacts which compromise the spectrum fitting, and removal of such noise is the first step before fitting. The spectrum is cleaned by identifying and removing the pixels (∼ 150) at the ends of the spectrum that contain noisy or unreliable data (infinite values). We calculate the mean and variance of the flux values in these regions and discard pixels that exceed a 3σ threshold.

---
[1] http://www-wfau.roe.ac.uk/6dFGS/xtalk/

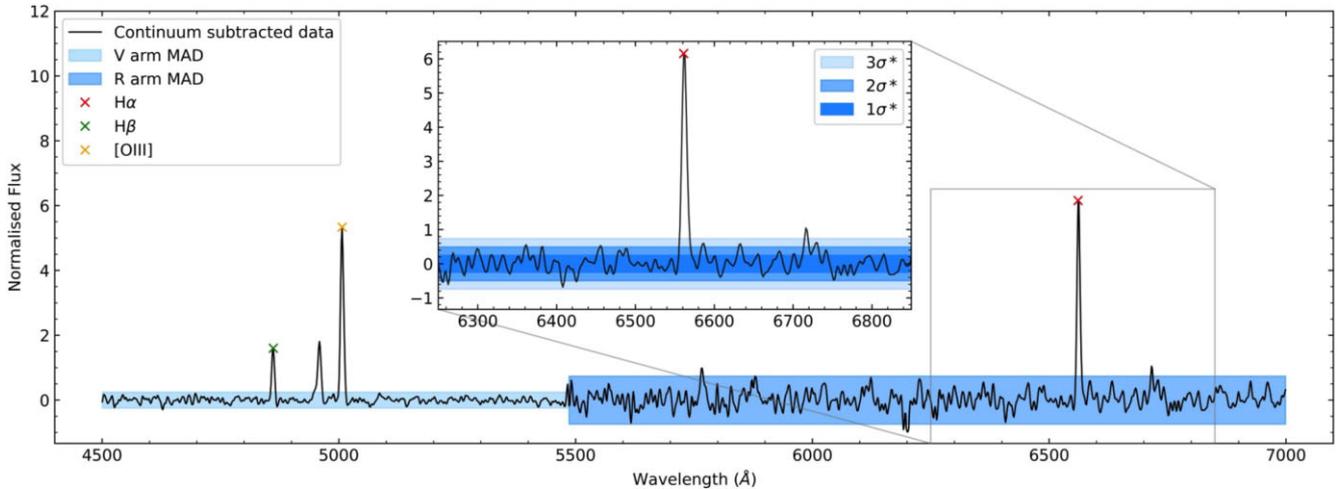

**Figure 4.** A continuum-subtracted example spectrum (Spec_ID 36658) shows the difference in the MAD values for the *V*-arm and *R*-arm. The light blue region indicates the MAD of the *V*-arm, while the darker shade corresponds to the MAD of the *R*-arm of the spectrum. The inset plot shows the zoomed-in view of the H$\alpha$ peak region along with the different MAD thresholds ($1\sigma^*$, $2\sigma^*$, and $3\sigma^*$).

We use a Continuum Fitting Tool (CFT) which employs Gaussian smoothing to separate emission lines from the continuum in a spectrum. The rationale behind this method is that emission lines are considered outliers to the mean noise level. Gaussian smoothing effectively spreads out these outliers while also reducing their peak intensity. By iteratively removing pixels that exceed a $3\sigma$ threshold above the Gaussian smoothed line, the spectrum gradually approaches the continuum profile. This process repeats until no pixels surpass the threshold, yielding a continuum-like profile. CFT systematically isolates the continuum, facilitating precise measurement and interpretation of emission features.

Following the cleaning procedures, the spectra are then corrected to the rest frame using the redshift data from the 6dFGS. A common reference frame helps in the analysis of the spectral features of the source. Rebinning involves re-sampling the spectrum on to the rest wavelength grid with uniform bin sizes, preserving the fluxes. This is achieved using the PYSYNPHOT python package (STScI Development Team 2013), which resamples the spectrum while keeping the fluxes intact. Every spectrum is also normalized to the mean of the continuum to homogenize the dataset.

### 3.2 Emission line detection – statistical approach

Emission lines can be differentiated from the noisy peak by statistical methods, that are more practical than visual inspection. We analyse the continuum subtracted spectrum statistically and employ the median absolution deviation (MAD) as the statistical measure to identify the emission lines in the spectrum. MAD is a measure of dispersion that assesses the variability or spread of a data set. It calculates the median of the absolute deviations of individual data points from the median of the data set. The MAD is often preferred over the standard deviation, especially in the presence of outliers or skewed data, as it is not influenced by extreme values. MAD reflects the typical deviation of data points from the median, and does not assume an underlying distribution of the data, making it suitable for non-normally distributed data sets. The following equation calculates the offset from the median:

$$\text{MAD} = \text{med}\left(|x_i - \text{med}(x)|\right), \tag{1}$$

where $x_i$ are the individual data values in the set, med($x$) denotes the median value of the data set. We use MAD as a measure to identify emission lines, by estimating the noise level. We exclude known outlier features such as the 5577Å Skyline, the *B*-band Telluric line at 6885 Å and prominent peaks, while calculating the MAD. This step ensures that we do not inflate the noise estimate for the good parts of the spectra. As described in Section 2, 6dFGS data has two separate arms, V and R; thus, we obtained MAD for the two arms separately using the 5577 Å Skyline as the reference point. Fig. 4 clearly shows the variation in MAD values for V and R arms, hereafter referred to as MAD$_V$ and MAD$_R$.

### 3.3 Automated peak identification

Before fitting the sources, we employ a method to identify emission line galaxies. With the MAD for the V arm and R arm, we now use the FIND_PEAKS SciPy python package (Virtanen et al. 2020) to confirm the presence of prominent peaks, that is [O III], H$\beta$, and H$\alpha$. The FIND_PEAKS Python package locates peaks within a designated wavelength range. The expected wavelength ranges of H$\beta$, H$\alpha$, and [O III] are given as arguments along with the MAD values. The MAD is scaled by 1.4826 to compare it with the standard deviation ($\sigma$) for a Gaussian distribution. Hereafter, we denote this robust estimate as $\sigma^*$, such that $n\sigma^*$ refers to $n \times 1.4826 \times$ MAD throughout this work.

Depending on the parent sample and the scientific objectives, emission-line selection criteria have varied over the years. Zhang (2023) uses S/N greater than 10 as a selection criteria, to collect all the low-redshift Type 2 AGN from the SDSS pipeline-classified main galaxies in DR16. This was adopted to ensure that the narrow emission lines, particularly the [N II] doublet and narrow H$\alpha$. While such stringent thresholds are appropriate for SDSS, which provides high spectral resolution and signal quality, our analysis is based on the 6dFGS data set, where the comparatively lower spectral resolution necessitates a different approach to defining reliable line detections. In the high redshift regime, Bongiorno et al. (2010) and Mignoli et al. (2019) identified 213 Type 2 AGN from the zCOSMOS survey, applying a 5$\sigma$ criterion for the [O III] line, while using a signal-to-noise ratio (S/N) > 2.5 for other emission lines. This 5$\sigma$ criterion is approximately equivalent to a 3$\sigma^*$ threshold ($\sigma^*$ $\approx 1.5\sigma$), which we tested on a random subset of 6dFGS spectra. It is important to note that our selection is based on the peak height identified using the FIND_PEAKS algorithm, rather than on fitted line parameters as adopted in the aforementioned studies. We find that the $3\sigma^*$ threshold provides a robust estimate for defining line peaks in our dataset. The inset plot in Fig. 4 illustrates $1\sigma^*$, $2\sigma^*$, and $3\sigma^*$ regions to aid visualization. From the figure, the estimation to use $3\sigma^*$ as the best threshold is justified as $1\sigma^*$ is below the noise level while $2\sigma^*$ is along the noisy peaks of the continuum. For H$\beta$ and [O III], MAD$_V$ is utilized, while MAD$_R$ is used for H$\alpha$. To ensure a robust AGN selection, we retain the sources whose [O III] and H$\alpha$ peak heights are greater than the $3\sigma^*$ threshold in the spectrum. Hence, 25 859 sources are selected out of 117 610 sources.

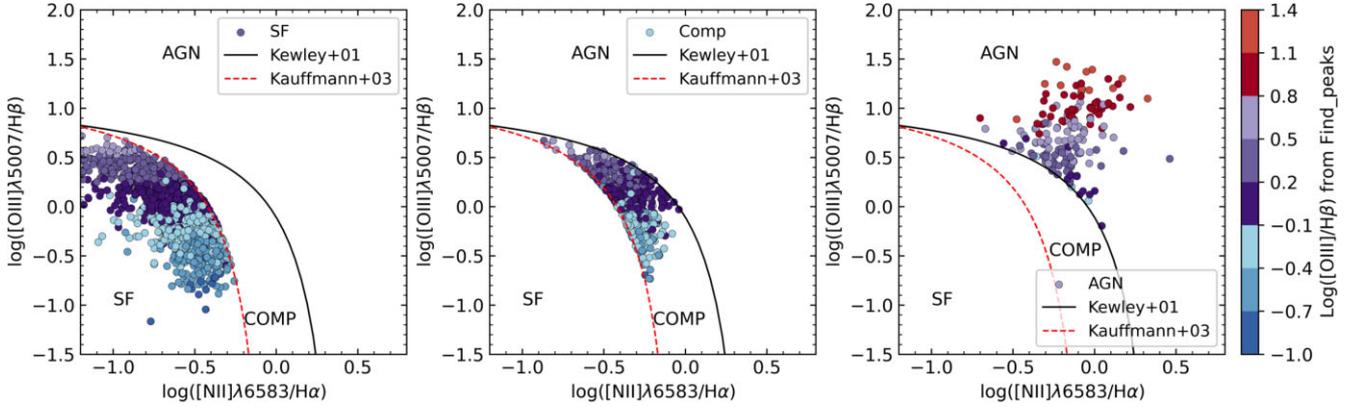

**Figure 5.** BPT diagram of 2553 sources. The left panel shows SF galaxies, colour-coded by log [O III]/H$\beta$ peak height measured with FIND_PEAKS. All SF sources in this random sample have peak heights < 0.8. The middle panel shows composites (COMP), displaying a similar trend. In contrast, AGN in the right panel have peak heights > −0.1, as expected. We adopt −0.1 as the threshold and remove SF sources with log [O III]/H$\beta$ < −0.1 from the sample.

### 3.4 Broad emission line galaxies removal

To remove Type 1 AGN, we cross-match our sample with H25, which is an extensive catalogue of Type 1 to 1.9 AGN in the 6dFGS sample. Type 1 AGN in H25 were identified using line fitting to verify the presence of broad lines and catalogued sources after a similar visual inspection of the spectra. The H25 sample contains all bright AGN from type 1 to type 1.5 and the majority of types 1.8 and 1.9, except for those with poor-quality spectra. Since each of the sources was line-fitted manually, there is no contamination from non-broad line sources. The cross-match excluded 993 Type 1 galaxies.

### 3.5 Sample filtration

Since the main focus of this work is to identify the Type 2 AGN, we streamline the process and develop a method that efficiently selects the AGN from our sample (reducing the count of SF galaxies). Before fitting galaxies from our sample, a random sample of ∼2500 galaxies that obey the $3\sigma^*$ criteria were selected. The peak height ratio of [O III] and H$\beta$ were computed across the sample using the FIND_PEAKS function. These sources were then fitted using PYQSOFIT to obtain the total flux values of the lines (further details about PYQSOFIT are in Section 3.6). Using the emission line ratios, we classify the sources into AGN, composites, and star forming (SF), based on the widely used diagnostic tool BPT diagram (a detailed explanation of BPT is in Section 4). Fig. 5 shows the classification in three subplots. The left panel shows the SF galaxies, colour-coded by log [O III]/H$\beta$ peak height measured with FIND_PEAKS. All SF sources in this random sample have peak heights < 0.8. Although the count is less, but composites (COMP) in the middle panel show a similar trend. In contrast, AGN in the right panel have a higher peak heights, i.e. above −0.1.

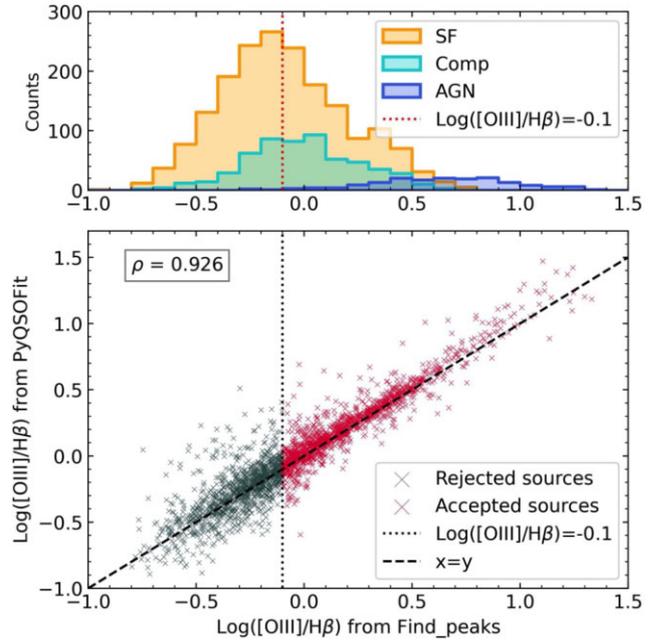

**Figure 6.** Bottom: Comparison between the [O III]/H$\beta$ peak height ratios derived using FIND_PEAKS and the flux ratios obtained from PYQSOFIT. Sources with log [O III]/H$\beta$ > −0.1 (shown in red) are selected, while those below this threshold (grey) are rejected. The black dotted line marks the selection threshold at log [O III]/H$\beta$ = −0.1, and the black dashed line represents the one-to-one correspondence ($x = y$). Top: Histogram showing the distribution of all sources coloured by their BPT classification: SF (orange), composites (turquoise), and AGN (blue). The red dotted line again indicates the log [O III]/H$\beta$ = −0.1 threshold.

The peak height ratio from FIND_PEAKS represents the height of the emission peak above the continuum level, rather than the integrated flux derived from spectral fitting. To assess whether this ratio can be used as a discriminating parameter, we compare the peak height ratio from FIND_PEAKS with the flux ratio obtained through the fitting procedure. The bottom panel of Fig. 6 shows the peak height ratio from FIND_PEAKS package and flux values of the sources measured from the fitting procedure i.e. PYQSOFIT. The figure demonstrates a linear correlation of the line ratio obtained by two different methods. This indicates that the peak height and the flux values are analogous, allowing us to perform a statistical analysis to exclude the SF galaxies from our sample. We further analyse our sample by plotting a histogram (the top panel of Fig. 6). The demographics reveal that AGN predominantly occupy the higher end of the histogram, as predicted in the BPT diagram (Fig. 5).

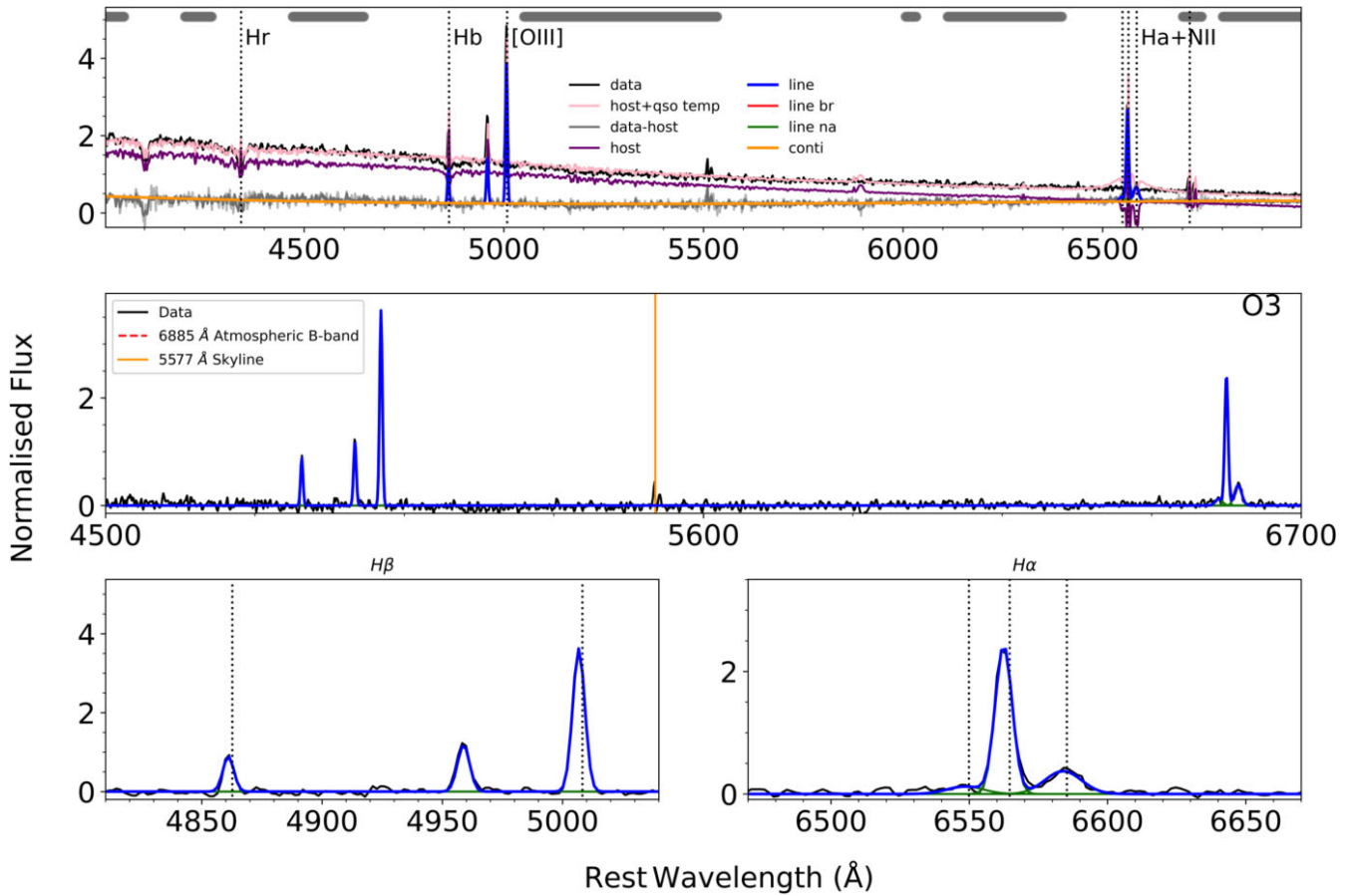

**Figure 7.** PYQSOFIT fitted spectrum for a source with Spec_ID 775. Top: 6dFGS raw spectrum (black), host-subtracted spectrum (grey), and the fitted continuum (orange). Middle: Continuum-subtracted spectrum (black) overlaid with the fitted emission lines (blue). Bottom: Zoom-in on the fitted H$\alpha$ and H$\beta$ regions. The spectrum has been corrected to the rest frame using the 6dFGS redshift. In this case, a slight redshift overestimation is visible but does not affect the fitting.

Based on this trend (Figs 5 and 6), we adopt −0.1 as the threshold and exclude all SF sources with log [O III]/H$\beta$ < −0.1 from the final sample. Consequently, the sample is partitioned into two groups. Sources with a peak height ratio exceeding −0.1 are earmarked for further investigation, while the remaining sources are slated for analysis in the latter part of the study. The proposed technique eliminates the need for visually inspecting the 8902 sources, thereby accelerating the overall process.

**3.6 Emission line fitting**

We used a modified version of PYQSOFIT[2] with an improvement to the functionality for grouping line parameters. This allows the fitting to accurately model the [O III]$\lambda\lambda$4959,5007 and the [N II]$\lambda\lambda$6548,6583 doublet. This version was also implemented using CFT as an alternative to the default template fitting for wavy or curved continuum. CFT has the advantage of being resilient to large-scale 6dFGS spectral artefacts and problems. We constrain the package to fit the narrow emission lines for the prominent peaks. Thus, the FWHM of the narrow lines: H$\alpha$, H$\beta$ and [N II] ≤ 700 km s$^{-1}$, while for [OIII] it is ≤ 1200 km s$^{-1}$. Some of the emission lines are grouped and fit simultaneously. H$\alpha$ and H$\beta$ are grouped and constrained to have similar shapes (FWHM, velocity offset and skew). The flux ratio [O III] and [N II] doublets are also fixed to 1/3. With the above initializations, the 15 964 sources are fitted using the modified PYQSOFIT code. Fig. 7 shows a sample fitting procedure using the PYQSOFIT code.

The sources are visually inspected to verify the fitting procedure. The need for eyeballing the fitting arises for various reasons. A few are highlighted:

(i) As the spectrum has two arms observed at two different points of time, there is a mismatch in the noise level. Also, this can lead to an inconsistent wavelength calibration, causing uncertain redshift calculation. No corrections have been applied to the tabulated 6dFGS redshifts used.

(ii) The continuum of the spectrum can be wavy or curved due to fibre transmission or data reduction issues, and then the usual continuum subtraction is not effective. Thus, we refit such sources by either fitting the *V*-arm and the *R*-arm separately or by using CFT.

---

[2] Found in https://github.com/JackHon55/PyQSOFit_SBL

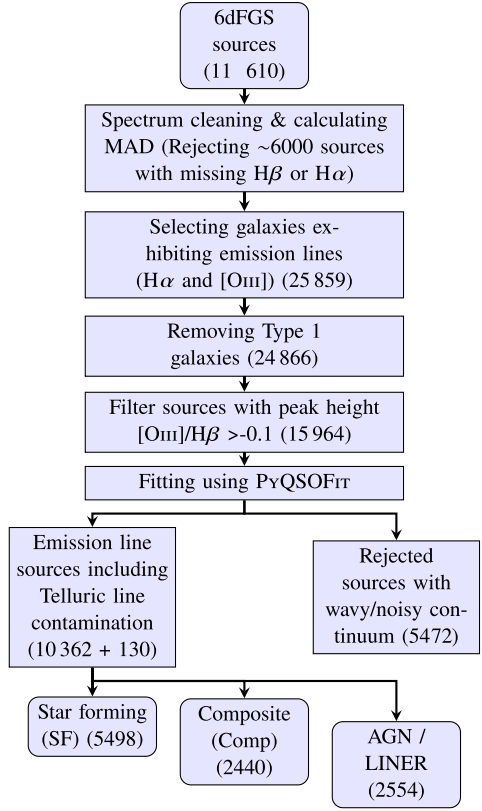

**Figure 8.** Flowchart summarizing the spectral pre-processing and emission line fitting steps, annotated with source counts retained at each stage.

(iii) As mentioned by J09, for some sources, the measured photon counts could be negative, thereby corrupting the emission line equivalent widths.

In these cases, we need to refit the continuum, to obtain proper fitting of emission lines. More details on the challenges of using 6dFGS spectra and the problems encountered are mentioned in detail in Section 2.3.

The flowchart in Fig. 8 shows the different steps of the spectrum pre-processing and fitting. A count of sources remaining after each process is also mentioned in the steps.

## 4 CLASSIFICATION OF SOURCES

The line ratio of [O III]/H$\beta$ serves as a critical parameter in discerning active galaxies, and was first proposed by BPT (Baldwin et al. 1981). The BPT diagram and its various classification schemes have become indispensable tools for understanding the ionization mechanisms and physical properties of galaxies. They have been extensively used in numerous studies exploring the nature and evolution of galaxies, such as those by Baldwin et al. 1981; Kewley et al. 2001; Kauffmann et al. 2003; Beckmann & Shrader 2012.

The BPT diagram originally compared the [O III]/H$\beta$ ratio to the [N II]/H$\alpha$ ratio to distinguish between different types of emission line galaxies. Exploring various pairs of ratios for AGN identification, (Veron 1981) introduced the metric [O III]/H$\beta$ versus [N II]/H$\alpha$, which offers advantages such as independence from exact reddening and accurate wavelength sensitivity calibration. Similarly, (Keel 1983) utilized three ratios – [N II]/H$\alpha$, [S II]$\lambda\lambda$6716,6731/H$\alpha$, and [O I]$\lambda$6300/H$\alpha$ – due to their robustness against reddening effects and wavelength sensitivity variations. Veilleux & Osterbrock (1987) conducted a comprehensive analysis, incorporating all three ratios as functions of [O III]/H$\beta$, further enhancing the understanding of AGN properties through multi-ratio investigations.

Using the BPT diagram, sources are classified into different regions based on their emission line ratios. SF galaxies typically occupy the lower-left region of the diagram, characterized by strong H$\alpha$ and [N II] emission lines relative to [O III] and H$\beta$. AGN and other ionized sources, such as LINERs (Low Ionization Nuclear Emission-line Regions), populate the upper-right portion of the diagram, exhibiting elevated [O III]/H$\beta$ and [N II]/H$\alpha$ ratios compared to SF galaxies.

Kewley proposed an empirical separation between SF galaxies and AGN/LINERs, defining a dividing line on the BPT diagram known as the 'Kewley' line (Kewley et al. 2001). This classification scheme provided a more robust and observationally motivated method for distinguishing between different types of emission line sources.

Kauffmann further refined the classification scheme by incorporating additional line ratios and introducing a theoretical demarcation, known as the 'Kauffmann' line. This line separates SF galaxies from those dominated by AGN activity based on the [O III]/H$\beta$ and [N II]/H$\alpha$ line ratios (Kauffmann et al. 2003).

We employ the same ratios in our work and classify our sources using the BPT diagram. The region between the 'Kewley line' and 'Kauffmann line' are called Composites (Comp). The following are the equations used to classify the galaxies into SF, composite, and AGN:

Star forming-
$$\log [\text{O III}]/\text{H}\beta = 1.3 + (0.61/(\log[\text{N II}]/\text{H}\alpha - 0.05)) \quad (2)$$

Composites-
$$\log [\text{O III}]/\text{H}\beta < 1.19 + (0.61/(\log[\text{N II}]/\text{H}\alpha - 0.47)), \quad (3)$$

$$\log [\text{O III}]/\text{H}\beta > 1.3 + (0.61/(\log[\text{N II}]/\text{H}\alpha - 0.05)) \quad (4)$$

AGN-
$$\log [\text{O III}]/\text{H}\beta > 1.19 + (0.61/(\log[\text{N II}]/\text{H}\alpha - 0.47)) \quad (5)$$

As detailed in Section 3.3, the selection criteria are based on the peak heights of [O III] and H$\alpha$ exceeding $3\sigma^*$, which results in some sources lacking strong H$\beta$ emission, since no threshold is imposed on that line. Approximately 1100 sources do not have a measurable H$\beta$ emission line. Excluding these sources would make the catalogue incomplete. Also, the low spectral resolution of the 6dFGS spectrum can lead to the H$\beta$ line being engulfed in noise. Keeping in mind the limitations, we include such sources, but flag them in the catalogue, so the user can exclude them according to their scientific requirements. Since the BPT diagram requires an H$\beta$ flux value, we assign a pseudo H$\beta$ flux, which is one-third of the H$\alpha$ flux (pseudo H$\beta$ values are not provided in the catalogue). The BPT diagram of 10 492 sources is as shown in Fig. 9. The left panel is the BPT diagram of $\sim$9400 sources with H$\beta$ emission line while the right panel shows the sources with pseudo H$\beta$ flux. Table 3 shows the count of AGN, SFs, and Comps identified using this classification.

Although the BPT diagram is a well-established method for identifying AGN, recent studies have pointed out its limitations. For instance, Agostino & Salim (2019) showed that a considerable number of AGN identified via X-rays reside in the SF region of the BPT diagram, indicating that the presence of star formation can mask the AGN signature in emission-line diagnostics. Nonetheless, the BPT diagram remains a valuable and widely used first-pass classification tool, particularly in large optical spectroscopic surveys.

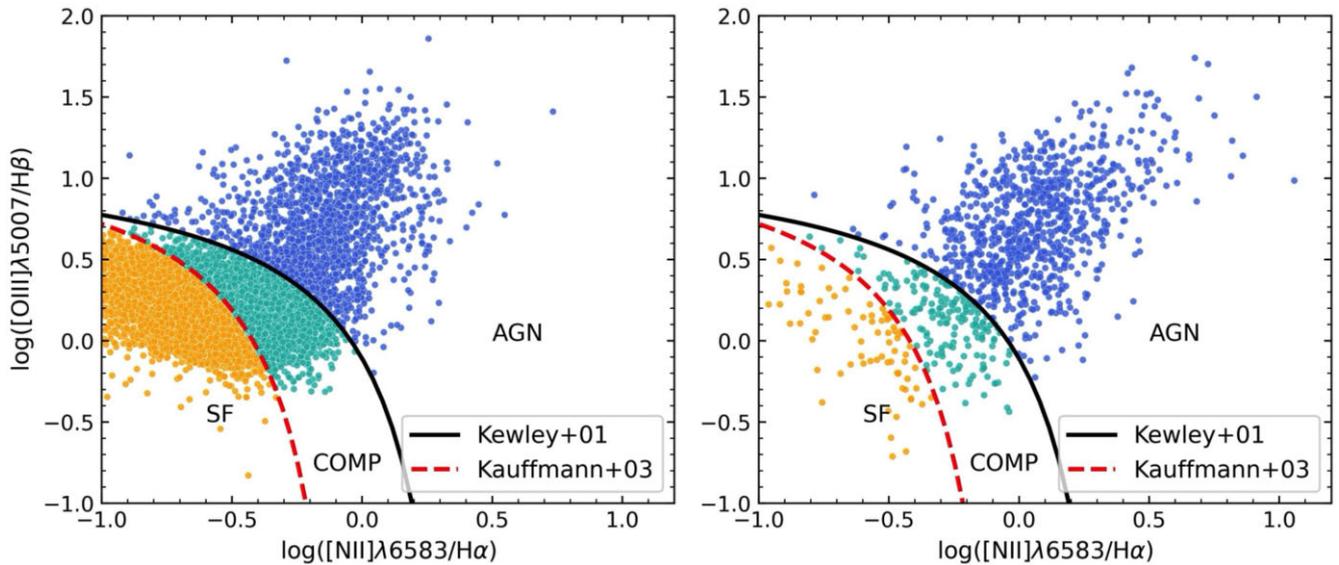

**Figure 9.** BPT diagram of all 10 492 sources in this work. Left: ∼9400 sources with detected H$\beta$ emission lines. Right: sources with pseudo-H$\beta$ flux.

**Table 3.** Classification of the sources.

| Classification | Count |
|---|---|
| SF | 5498 |
| Comp | 2440 |
| AGN | 2554 |

## 5 PROPERTIES OF THE CATALOGUE

### 5.1 Photometric flux calibration and luminosity estimation

Since optical spectra of 6dFGS are not flux calibrated, we do so using the SkyMapper Southern Survey DR4 (SMSS, Wolf et al. 2018; Onken et al. 2024). We cross-match our 10 492 sources with the SMSS catalogue and find 10 351 sources with SMSS detection within 3 arcsec. This ensures that the photometry data used for calibration is of the same source. As a quality check, we only attempt to calibrate the sources for which the FLAGS column of the SMSS data is less than 4 and make sure the photometry data is available for both *g* and *r* filters. To ensure robust photometry, we require the source to have been observed by SkyMapper more than 5 times. Thus, the calibrated flux values are only provided for 9549 sources. The remaining sources are included in the catalogue with the Flux counts instead of calibrated flux values.

We calibrate the flux by following the methods used in H25, where the emission line flux of [OIII] and H$\beta$ is calibrated using:

$$F_\lambda(x) = \frac{Counts(x)}{Ctm(x)} f_p\lambda(ctm(x)) = EW(x) f_p\lambda(ctm(x)) \quad (7)$$

Here, we determine the $F_\lambda(x)$ which depends on the counts ($counts(x)$), continuum value near the emission line ($ctm(x)$) and the flux values at the particular wavelength ($f_p\lambda(ctm(x))$). The first term in the right-hand side equation is the Equivalent Width (EW) of the emission line, which is provided in the catalogue as the uncalibrated flux (FLUX_X, see Section 5.3). To match the 6dFGS fibre diameter of 6.7 arcsec, we adopt the 6 arcsec SMSS photometric aperture and flux-calibrate the V spectra from *g* ($\lambda_{eff}$ = 5100 Å) and the R spectrum from *r* ($\lambda_{eff}$ = 6170 Å). The magnitude values at 6 arcsec along with the calculated photometric flux for both the bands are provided in the catalogue (refer Table 4). The photometric flux FLUX_G and FLUX_R represents fluxes at effective wavelength 5100 and 6170 Å, respectively (Bessell et al. 2011; Tonry et al. 2018).

The integrated flux is de-reddened using $E(B-V)$ values from the GNILC dust maps, provided by Planck 2018 (Planck Collaboration IV 2020) with Fitzpatrick law (Fitzpatrick 1999) and $R_V = 3.1$. The de-reddened fluxes are then converted to luminosities. The catalogue provides the calibrated flux value of [O III] and H$\beta$ along with the luminosity. The column CFLUX_HB and CFLUX_OIII are the calibrated flux obtained by this method for [O III] and H$\beta$. The de-reddened flux of these two emission lines are also provided in the catalogue as the columns DEFLUX_HB and DEFLUX_OIII. The log luminosities calculated from the de-reddened fluxes are included as LOGL_HB and LOGL_OIII. The corresponding errors for the calibrated flux and the log luminosity for [O III] and H$\beta$ have the _ERR suffix. For the sources with Pesudo H$\beta$, we only provide the calibrated flux and luminosity for [O III].

### 5.2 Source detection algorithm

Within the 6.7 arcsec diameter of 6dFGS fibre, there exist cases where the galaxy is contaminated by another source. Photometry values from such blended sources cannot be trusted to be the 6dFGS source. We flag such sources in the catalogue as dubious calibration.

Although these blended sources can often be identified using data from the *Gaia* satellite (Gaia Collaboration 2016, 2023), not all of our sources have corresponding *Gaia* data. SMSS provides the count of *Gaia* sources within a 15 arcsec radius; however, the 6dFGS fibre diameter is less than half the distance. It is also possible for both *Gaia* and SkyMapper to detect bright H II regions of a galaxy as neighbouring sources.

Apart from the existing catalogued data, we can look for the blended sources visually using the PanSTARRS and DSS2 images (Chambers et al. 2016). To address this issue, we divide our calibrated 9549 sources based on their declination. If the sources lie within the PanSTARRS observing limit (0° to −30°), then we obtain the images in gri filters (5218 sources). DSS2 Images in blue, red, and IR filters are used for the rest of the 4331 sources. The three filters are used to make a single RGB colour image of every source. We examine every image and make sure that there is only one source within the 6.7 arcsec diameter. The expected contaminants include (a) a star nearby, (b) another companion or a satellite galaxy, and (c) a merger or interaction. The column IM_FLAG in the catalogue indicates such sources where the value 0 indicates a single object, 1 corresponds to contamination, and all uncalibrated sources are given a value 2.

**Table 4.** The catalogue table description. † see description in text.

| Column | Type, Unit | Description |
| --- | --- | --- |
| SPEC_ID | int | Spec ID of the 6dFGS source as mentioned in dataset |
| DUPLICATE_ID | string | Duplicate Spec ID of the sources (separated by ';') |
| RA | int, Degrees | Right ascension |
| DEC | int, Degrees | Declination |
| 6dFGS_NAME | string | 6dFGS Name |
| PROG_ID | int | 6dFGS identifier, see Table 1 |
| QUALITY | int | Quality flag |
| Z | float | Redshift |
| FWHM_X | float, km s$^{-1}$ | Full width Half Maximum of X† |
| FWHM_X_ERR | float, km s$^{-1}$ | Corresponding error |
| PEAK_X | float, Å | Peak wavelength of X† |
| PEAK_X_ERR | float, Å | Corresponding error |
| FLUX_X | float, Counts | Flux of X† |
| FLUX_X_ERR | float, Counts | Corresponding error |
| LOG_OIII_HB | float | Log of ratio of [O III] and H$\beta$ |
| LOG_NII_HA | float | Log of ratio of [N II] and H$\alpha$ |
| BPT | string | BPT classification of the sources. Asterisk (∗) indicates absence of H$\beta$ emission line |
| FLAG | int | Contaminated by Crosstalk (1), Telluric line (2), no contamination (0) |
| SMSS_ID | int | SkyMapper unique identifier |
| SMSS_DIST | float | Distance of the Closest SkyMapper target |
| R_MAG | float, AB mag | SkyMapper *R*-band magnitude at 6 arcsec aperture |
| R_MAG_ERR | float, AB mag | Corresponding error |
| G_MAG | float, AB mag | SkyMapper *G*-band magnitude at 6 arcsec aperture |
| G_MAG_ERR | float, AB mag | Corresponding error |
| EBMV | float | $E(B-V)$ by Planck Collaboration IV (2020) |
| FLUX_R | float, erg s$^{-1}$ cm$^{-2}$ Å$^{-1}$ | SkyMapper photometry constructed *R*-band flux |
| FLUX_G | float, erg s$^{-1}$ cm$^{-2}$ Å$^{-1}$ | SkyMapper photometry constructed *G*-band flux |
| CFLUX_HB | float, erg s$^{-1}$ cm$^{-2}$ | Photometry estimated integrated flux of H$\beta$ |
| CFLUX_HB_ERR | float, erg s$^{-1}$ cm$^{-2}$ | Corresponding error |
| DEFLUX_HB | float, erg s$^{-1}$ cm$^{-2}$ | Corresponding de-reddened flux |
| CFLUX_OIII | float, erg s$^{-1}$ cm$^{-2}$ | Photometry estimated integrated flux of [O III] |
| CFLUX_OIII_ERR | float, erg s$^{-1}$ cm$^{-2}$ | Corresponding error |
| DEFLUX_OIII | float, erg s$^{-1}$ cm$^{-2}$ | Corresponding de-reddened flux |
| LOGL_HB | float, erg s$^{-1}$ | Log H$\beta$ luminosity derived from de-reddened estimated flux |
| LOGL_HB_ERR | float, erg s$^{-1}$ | Corresponding error |
| LOGL_OIII | float, erg s$^{-1}$ | Log [O III] luminosity derived from de-reddened estimated flux |
| LOGL_OIII_ERR | float, erg s$^{-1}$ | Corresponding error |
| IM_FLAG | int | Single sources (0), more than one object (1), uncalibrated sources (2) |
| 2MASS_ID | string | 2MASS unique identifier |
| JMAG | float, mag | 2MASS *J*-band magnitude from point source catalogue |
| JMAG_ERR | float, mag | Corresponding error |
| HMAG | float, mag | 2MASS *H*-band magnitude from point source catalogue |
| HMAG_ERR | float, mag | Corresponding error |
| KMAG | float, mag | 2MASS *K*-band magnitude from point source catalogue |
| KMAG_ERR | float, mag | Corresponding error |
| JEXT | float, mag | 2MASS *J*-band total magnitude from extended source catalogue |
| JEXT_ERR | float, mag | Corresponding error |
| HEXT | float, mag | 2MASS *H*-band total magnitude from extended source catalogue |
| HEXT_ERR | float, mag | Corresponding error |
| KEXT | float, mag | 2MASS *K*-band total magnitude from extended source catalogue |
| KEXT_ERR | float, mag | Corresponding error |

## 5.3 Columns description

The catalogue comprises all the emission line sources, fitted using PYQSOFIT. Sources potentially affected by 6885 Å B-band telluric line contamination and sources that have spectra that were compromised during the observations are flagged, as new spectra may enable the completion of the sample. Sources missing H$\beta$ emission lines are also flagged, but included in the catalogue. Thus, the catalogue comprises 10 492 sources, each characterized by various properties delineated in different columns. The catalogue contains different properties of the sources, described below:

(i) **Source properties:** The positional data such as RA, DEC of the sources are obtained from the 6dFGS spectral data. There can be differences between the positions obtained from the published 6dFGS catalogue and the observed positions of the sources reported in the spectra (used in this work). Redshift (Z) recorded in the survey is taken as one of the inputs for fitting, thus added to the catalogue.

(ii) **6dFGS Data:** Some important data like 6DFGS_NAME, SPEC_ID, QUALITY are obtained directly from the headers of the 6dFGS fits file. Some of the sources have been re-observed due to poor weather conditions or technical issues, thus 6dFGS have multiple spectra for a single source. These sources are fitted and the best among them is selected for our catalogue. The duplicate IDs of such sources are indicated in the column DUPLICATE_ID.

(iii) **Fitting Results:** The FWHM, Peak wavelength, and uncalibrated flux of the emission lines are tabulated along with the errors. The column names are defined as FWHM_X, PEAK_X, and FLUX_X where X denotes the [O III], H$\beta$, H$\alpha$, and [N II] emission lines. Their corresponding errors are provided with the suffix _ERR.

(iv) **BPT classification:** SF, composite, and AGN classification. An asterisk indicates the absence of H$\beta$.

(v) **Contamination flag:** We flag the sources affected by crosstalk and the telluric line at 6885 Å. Thus, a flag is added to the column such that, 0 – no contamination, 1 – cross-talk, 2 – telluric.

(vi) **Flux calibration:** We also calibrate a part of the sources based on SkyMapper photometry (more about this in Section 5.1). The catalogue provides the associated SkyMapper Southern Survey DR4 object_id and the angular offset. We also provide the photometric data used for calibration.

(vii) **Image flag:** Apart from the contamination flags, which are based on the visual inspection of the spectrum, we also add another flag based on the optical image of the source. Section 5.2 explains the need for this flag along with the description of the usage.

Table 4 contains the column name with a short description of the column. The catalogue collectively forms a comprehensive resource for studying the properties and characteristics of galaxies within the 6dFGS data set, facilitating various new scientific analyses. The final catalogue is uploaded to CDS.

## 6 DISCUSSION

### 6.1 Comparison with other catalogues

A limited number of Southern Sky AGN catalogues have been published with the most similar work reported in C22, where 12 156 narrow-line AGNs were identified from the 6dFGS sample, by computing the S/N for the V arm and R arm separately. The flux has been obtained for sources that have S/N $\geq$ 2.0. However, the non-uniform noise across the two arms of 6dFGS spectrum and the presence of spurious features can overestimate or underestimate the S/N ratio. Therefore, in some cases the emission line flux appears significant but the automated S/N is affected by local artefacts, which leads to a reduced S/N ratio. Thus, in this work, we do not rely on the S/N of the spectrum but rather visually inspect the noise level and the fitting. We rely on our selection based on FIND_PEAKS package to identify the emission line, with $3\sigma^*$ MAD as a threshold to compute the strength of the line. To fit the emission lines, we use a modified PYTHON package, which includes the main routine – Fe II templates, an input line-fitting parameter list, host galaxy templates, and a dust reddening map to extract spectral measurements from the raw fits. This refined fitting routine provides us the spectral measurements with increased accuracy.

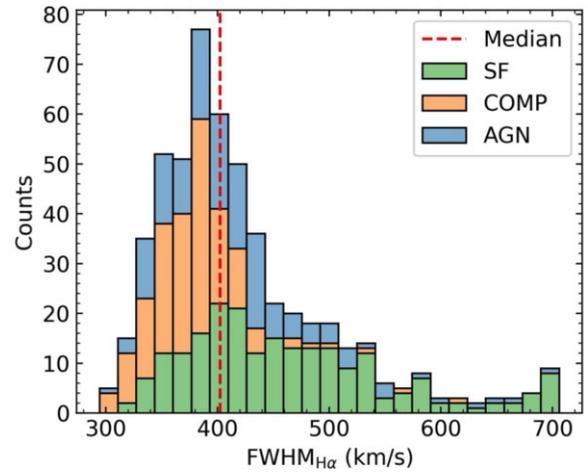

**Figure 10.** Histogram showing the distribution of FWHM$_{H\alpha}$ for sources classified as Type 1 in C22, with FWHM values derived using PYQSOFIT. The histogram is colour-coded by BPT classification: AGN (blue), SF (green), and composite (orange). The red dashed line marks the median value at 402.3 km s$^{-1}$.

3846 sources have been matched with C22, while 6646 new sources have been identified (from which, 1056 are AGN and 1058 Composite). The matched sources are further analysed to check for variation. 614 NLGs have been classified as Type 1 in C22, among which 286 are AGN using the Kewley et al. (2001) criteria. The FWHM at H$\alpha$ is found to be less than 700 km s$^{-1}$ from our fitting while the acceptable range of FWHM for Type 2 is $\gtrsim$1000 km s$^{-1}$. The Fig. 10 shows the distribution of FWHM at H$\alpha$ with median value 402.3 km s$^{-1}$. Similarly, 36 sources classified as Type 2 were found in H25.

Fig. 11 shows the flux ratios computed in our catalogue compared to the flux ratios in C22. We see the scatter is centred around the $x = y$ line, indicating a linear relation. The RMS value for the log([O III]/H$\beta$) is 0.19, while the RMS for log([N II]/H$\alpha$) is 0.14. A subset of sources ($\sim$9000) which are rejected by our method but present in C22 were selected and analysed in detail. 102 sources do not satisfy the redshift limit, indicating missing prominent emission lines. Around 6000 sources that have MAD values less than the threshold chosen, indicating a low S/N.

Prior to C22, Z19 has also identified AGN using the 6dFGS sample. Zaw has uniformly selected an All-sky catalogue of 6562 optical narrow-line AGN. The method deployed in that work is the same as C22, that is, to identify emission line galaxies based on S/N. As mentioned earlier, a more robust statistical approach to identifying the emission line galaxies has been utilized in our work.

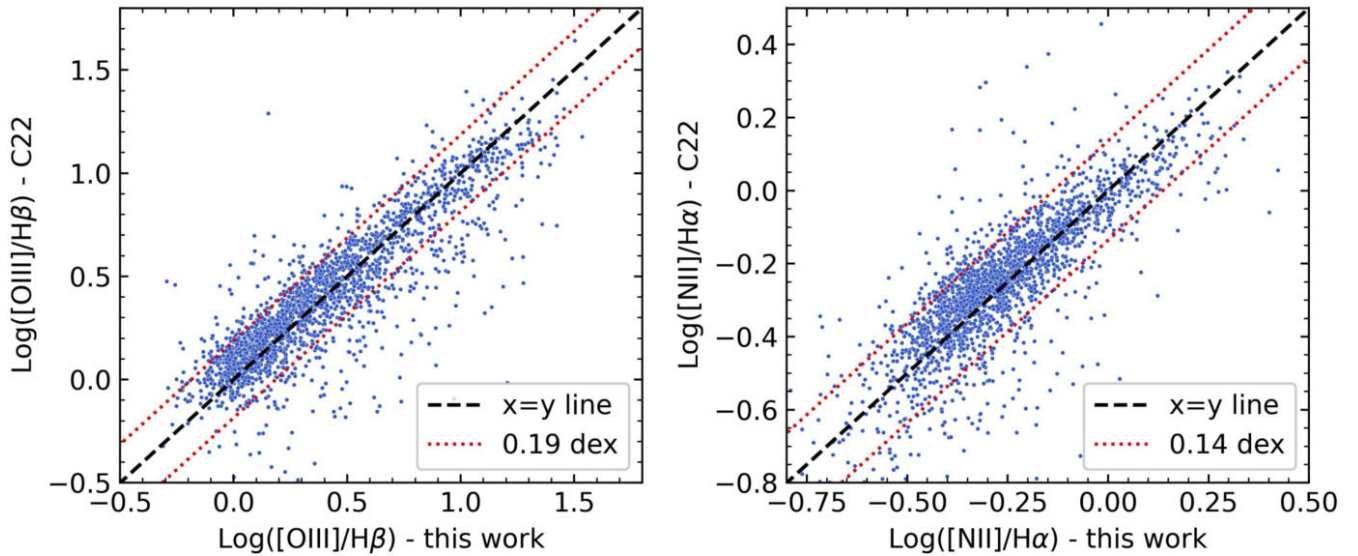

**Figure 11.** Comparison of flux ratios derived in this work with those from C22. Left: log([O III]/H $\beta$). Right: log([N II]/H $\alpha$). The *x*-axis shows values from this study, while the *y*-axis shows those from C22. Dotted lines (red) indicate the RMS limits.

### 6.2 Characteristics of AGN and composite

As mentioned in Section 2 and Table 1, the whole 6dFGS sample is divided into a main flux-limited sample and a set of auxiliary targets. These targets were added to the main survey from different multiwavelength catalogues. The PROG_ID column gives this information, where sources with PROG_ID < 10 are from the main target sample while sources with PROG_ID > 10 are auxiliary targets. In this subsection, we try to analyse the combined AGN and Composite (hereafter, AC) sample in the catalogue based on the nature of the target selection. We compare the redshift, 2MASS $K$ magnitude (and $K$-band extended magnitude) of the sample with emission line fluxes and their ratios. For accuracy, we use only the calibrated sources for which H $\beta$ emission line flux is observed. The $J$-, $H$-, and $K$-band magnitudes from both 2MASS point source catalogue and the 2MASS extended catalogue (XSC) is provided in the catalogue.

Fig. 12 gives a detailed analysis of the sources. As mentioned in the Section 3.3, we select the sources with H $\alpha$ and [O III] emission lines which restrict the redshift range to $z < 0.26$. We find that the auxiliary ACs (blue) have a higher redshift with a median value of 0.07 while the median of the main target (orange) is found to be 0.03 (see Fig. 12a). The main sample consists of much brighter sources in lower redshifts (less than $z \sim 0.14$) while the auxiliary targets are dimmer and have a redshift range of up to our limit of 0.26. The two subplots, b and c, in Fig. 12 study this luminosity difference in detail. Fig. 12(b) indicates that there is not much difference in the log [O III] line luminosity for these two sets of targets, and a similar result is shown in Fig. 12(c). The log [O III]/H $\beta$ line ratio is computed from the calibrated flux of the corresponding emission line. Most of the ACs in the auxiliary sample have higher luminosity, but this could be associated with the smaller number of auxiliary sources compared to the main targets. The ACs in the auxiliary are much dimmer than the main targets as seen further in plots 12(d), 12(f), and 12(h), where we compare the flux ratio and the luminosity with the $K$-band magnitude. The median of 2MASS $K$ band magnitude for the Auxiliary sources is 14.03 while for main targets it is 13.56. From the BPT diagram (see Fig. 9), we see that the pure AGN have log [O III]/H $\beta$ more than $\sim 0.8$, that is, no SF or Composite galaxy have the ratio more than 0.8. In the plots 12(d) and 12(e), we see that most of the pure AGN in our catalogue are from the main targets, as indicated by the dotted line.

We cross-match our sources with the 2MASS XSC catalogue and compare the $K_{ext}$ magnitude with the point source K magnitude. Thus, $K - K_{ext}$ gives the information about the apparent size of the source. This is only used as a proxy and further study on morphology and other galaxy properties is needed to confirm the results. From the plots 12(e), 12(g), and 12(i), we see that the auxiliary ACs are point-like objects and extended sources in the main target sample. The plot 12(d) shows that there less a number of strong, point-sourced AGN in the auxiliary sample while most of the extended pure AGN are in the main target sample. Thus, the larger the apparent size, the weaker and dimmer the AGN type. In Figs 12(g) and 12(i), we observe a trend in the sources. The more the apparent size, the dimmer it gets, irrespective of auxiliary or main targets. A similar trend was also observed in H25.

### 6.3 Rejection criteria and caveats

Approximately 5400 sources are rejected after visually inspecting the spectrum and are excluded from this catalogue. They presented a number of spectroscopic data or analysis problems:

(i) Almost 12 per cent Sources with a bad, wavy and noisy spectrum whose continuum is almost sinusoidal. Some of the bad spectra are shown in Fig. 13.

(ii) The 5577 Å skyline falling on H $\beta$ or [O III], which results in inaccurate line flux measurements. We remove 593 such sources ($\sim 11$ per cent) from our catalogue.

(iii) As mentioned in Section 5.3, around 4 per cent of the sources are observed more than once. We reject the source, but flag their Spec_ID in the duplicate ID column of the catalogue.

(iv) Approximately, 2 per cent of the sources have noisy peaks at the rest wavelength, that are not emission lines. Around 1200 sources ($\sim 22$ per cent) are missing at least one of the three emission lines ([O III], H $\alpha$, or [N II]).

(v) In sources for which the [O III] emission line is just above the $3\sigma^*$ MAD threshold, the [O III]$\lambda 4959$ emission line is often hidden by noise. PYQSOFIT identifies the [O III] line based on the [O III]$\lambda 4959$, absence of latter alters the Gaussian profile of [O III] and also [N II]. Around 1200 sources are rejected from the catalogue as the presence of three emission lines are crucial for classification. Combination of weak [O III]λ4959 emission and noisy peaks around [O III] leads to ambiguous selection, are removed from the catalogue (25 per cent).

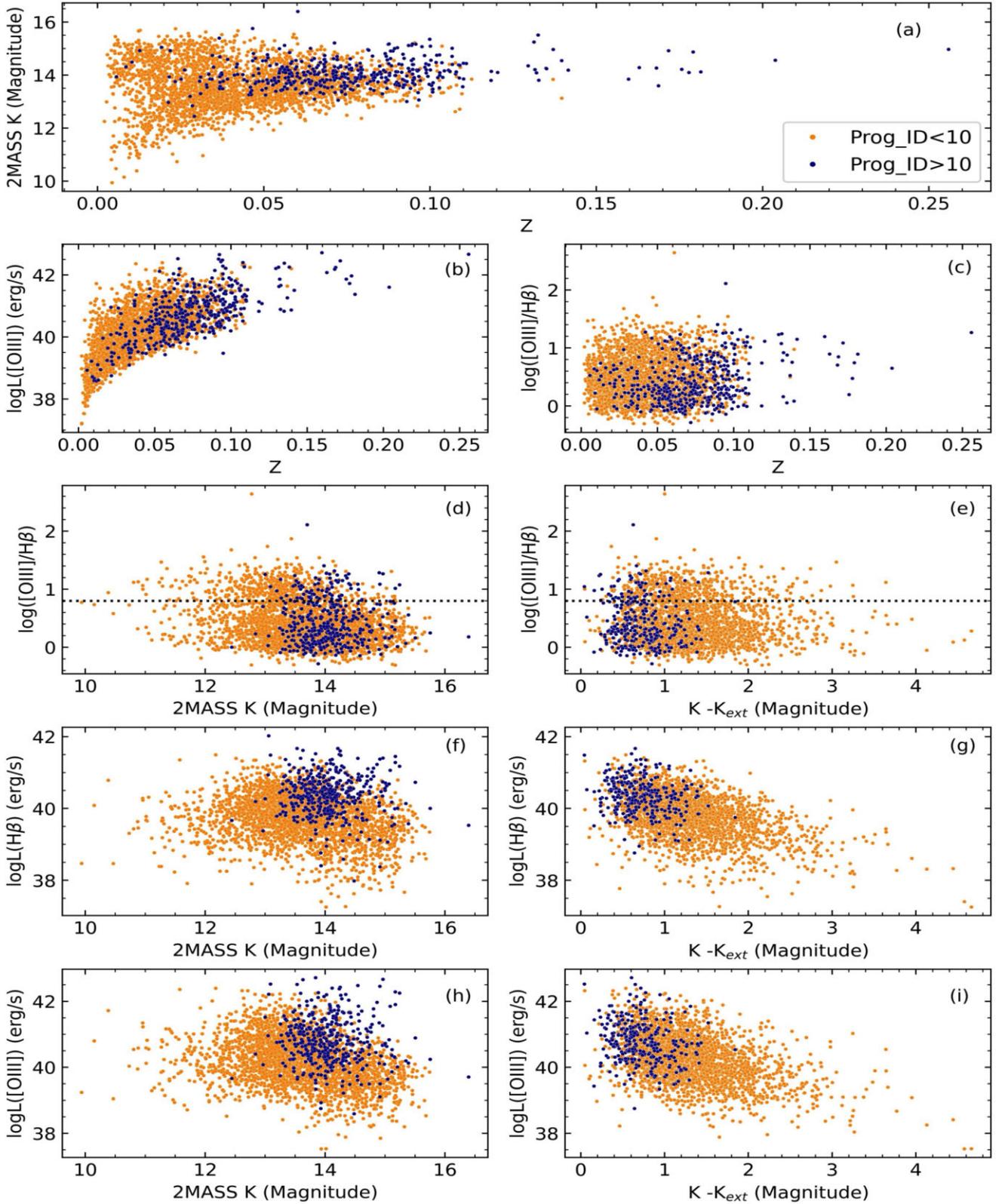

**Figure 12.** The figure compares calibrated sources from the auxiliary (PROG_ID > 10, in blue) and main target (PROG_ID < 10, in orange) samples. Panel (a) compares the 2MASS $K$-band magnitude with redshift. Panels (b) and (c) show comparisons of the [O III] log (luminosity) and the log([O III]/H$\beta$) emission-line ratio, respectively. Panels (d) and (e) plot the line ratio against the 2MASS $K$-band magnitude and the $K - K_{\text{ext}}$ magnitude. The dotted line in each plot indicates the lower limit of the line ratio, above which pure AGN are expected. Panels (f) and (h) compare the H$\beta$ and [O III] log luminosities with the 2MASS $K$-band magnitude. Panels (g) and (i) compare the same emission-line luminosities with the $K - K_{\text{ext}}$ magnitude.

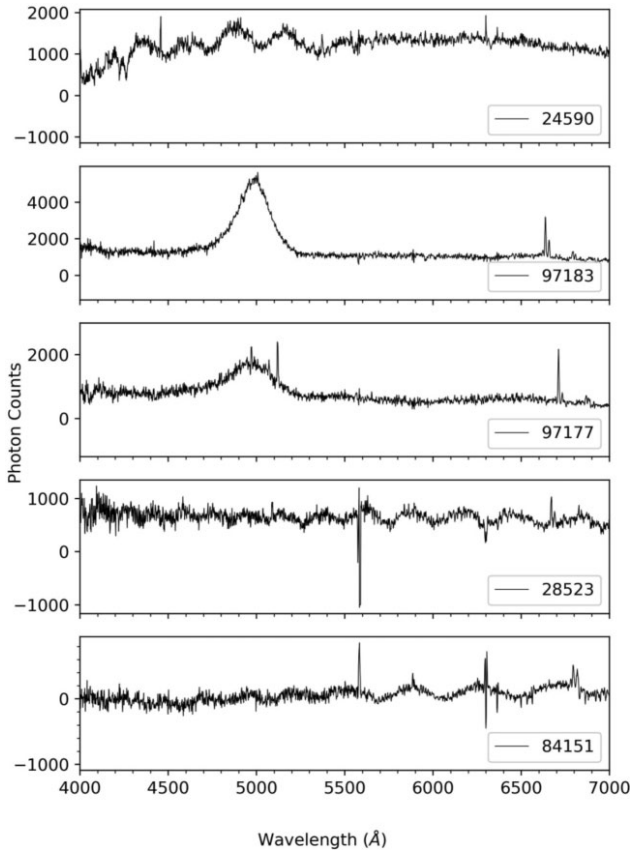

**Figure 13.** Figure showing the selected examples of problematic spectra which are wavy, or noisy. Their Spec_IDs are mentioned in legend.

## 7 CONCLUSION

In this work, we identified Type 2 AGN sources in the 6dF Galaxy survey (6dFGS) by visually inspecting the spectra after applying an emission-line fitting method. Before fitting, spectra were pre-processed, which included cleaning the spectrum by removing unreliable pixels. The continuum-subtracted spectrum was then used to identify the emission line galaxies by employing the MAD as a statistical cut. Broad emission line galaxies were removed from the sample by cross-matching with H25.

We employed the PYQSOFIT package to fit the key emission lines to the continuum-subtracted and cleaned spectra. The resultant fitting values include the flux counts, FWHM, and Peak of the four prominent emission lines (H$\beta$, [O III], H$\alpha$, and [N II]). The emission line flux obtained from the package was used to plot the BPT diagram to classify narrow emission line galaxies into SF, composite and AGN sources.

Starting from the parent sample of 136 304 spectra, we select 24 866 narrow emission line galaxies. Compared to other studies that identify Type 2 AGN (Bongiorno et al. 2010; Mignoli et al. 2019; Zhang 2023, C22, Z19), our approach stands out for its emphasis on visual verification of statistically selected sources, resulting in a more refine catalogue. The use of the $3\sigma^*$ criterion (for [O III] and H$\alpha$) for selecting emission-line galaxies, rather than relying on a S/N ratio, proves effective for 6dFGS spectrum, as the MAD is more robust to outliers and provides a more reliable estimate of the local noise level. The spectra of the selected galaxies are then visually inspected to increase the accuracy of the fitting measurements. Any correction needed in the fitting procedure has been manually done to reduce contamination. This step ensures the credibility of the fitting procedure, giving us the accurate flux values and removing the contamination (flagged *B*-band Absorption line at 6885 Å). The cross-talks mentioned in Section 2.3 are noted and removed wherever needed. 10 492 Narrow emission line galaxies have been identified in the 6dFGS data, which includes 2554 AGN and 2440 composites.

Using the SkyMapper Photometry data, we have calibrated 9549 sources based on the flags mentioned in the SMSS paper. The calibrated sources are then visually inspected using the PanSTARRS and DSS images to verify the presence of a single source within 6dFGS's 6.7 arcsec aperture. We classify Type 2 AGN based on emission line ratios, which requires the emission to originate from a single host galaxy. A total of 8848 isolated sources were identified that show no signs of contamination from mergers, interacting galaxies, nearby stars, or other companions. Sources with more than one galaxy within the aperture limit are flagged, as these may be useful for future studies exploring the connection between AGN activity and galaxy interactions. Researchers interested in the role of host galaxy morphology or AGN triggering in merging systems can make use of these flagged subsamples.

The catalogue offers a rich data set for diverse astrophysical investigations. By utilizing the PYQSOFIT derived parameters and BPT classifications, we intend to identify and study Changing Look AGN using the upcoming LSST. By tracking changes in emission line strengths and continuum properties, we can gain insights into the physical processes driving these variations, enhancing our understanding of the AGN.

## ACKNOWLEDGEMENTS

We thank the anonymous referee for their generous and constructive comments in finalizing the manuscript. This research was supported [or 'Parts of this research were supported'] by the Australian Research Council Centre of Excellence for All Sky Astrophysics in 3 Dimensions (ASTRO 3D), through project number CE170100013. The national facility capability for SkyMapper has been funded through ARC LIEF grant LE130100104 from the Australian Research Council, awarded to the University of Sydney, the Australian National University, Swinburne University of Technology, the University of Queensland, the University of Western Australia, the University of Melbourne, Curtin University of Technology, Monash University and the Australian Astronomical Observatory. SkyMapper is owned and operated by The Australian National University's Research School of Astronomy and Astrophysics. The survey data were processed and provided by the SkyMapper Team at ANU. The SkyMapper node of the All-Sky Virtual Observatory (ASVO) is hosted at the National Computational Infrastructure (NCI). Development and support of the SkyMapper node of the ASVO has been funded in part by Astronomy Australia Limited (AAL) and the Australian Government through the Commonwealth's Education Investment Fund (EIF) and National Collaborative Research Infrastructure Strategy (NCRIS), particularly the National eResearch Collaboration Tools and Resources (NeCTAR) and the Australian National Data Service Projects (ANDS). This work has made use of data from the Pan-STARRS1 Surveys (PS1) and the PS1 public science archive have been made possible through contributions by the Institute for Astronomy, the University of Hawaii, the Pan-STARRS Project Office, the Max-Planck Society and its participating institutes, the Max Planck Institute for Astronomy, Heidelberg and the Max Planck Institute for Extraterrestrial Physics, Garching, The Johns Hopkins University, Durham University, the University of Edinburgh, the Queen's University Belfast, the Harvard-Smithsonian Center for Astrophysics, the Las Cumbres Observatory Global Telescope Network Incorporated, the National Central University of Taiwan, the Space Telescope Science Institute, the National Aeronautics and Space Administration under Grant No. NNX08AR22G issued through the Planetary Science Division of the NASA Science Mission Directorate, the National Science Foundation Grant No. AST-1238877, the University of Maryland, Eotvos Lorand University (ELTE), the Los Alamos National Laboratory, and the Gordon and Betty Moore Foundation. This work has made use of data from The Digitised Sky Survey, which was produced at the Space Telescope Science Institute under U.S. Government grant NAG W-2166. The images of these surveys are based on photographic data obtained using the Oschin Schmidt Telescope on Palomar Mountain and the UK Schmidt Telescope. The plates were processed into the present compressed digital form with the permission of these institutions.

**DATA AVAILABILITY**

6dFGS data can be downloaded from their official website – http://www-wfau.roe.ac.uk/6dFGS/. Pan-STARRS1 image metadata were retrieved using the public ps1filenames.py service hosted by the Mikulski Archive for Space Telescopes (MAST): https://ps1images.stsci.edu. DSS images were obtained from the *SkyView* Virtual Observatory service maintained by NASA's Goddard Space Flight Center (https://skyview.gsfc.nasa.gov). Both the data are publicly available and were accessed using ASTROPY python package. The Type 2 AGN catalogue produced from this work will be available at CDS via anonymous ftp to cdsarc.u-strasbg.fr (130.79.128.5) or via https://cdsarc.cds.unistra.fr/viz-bin/cat/J/MNRAS.

This paper has been typeset from a T<sub>E</sub>X/L<sup>A</sup>T<sub>E</sub>X file prepared by the author.